\documentclass[aps,
              pra,
               twocolumn,
               showpacs,
               amsmath,
               amssymb,
               superscriptaddress,
               reprint,
               10pt
              ]{revtex4-1}

\usepackage[hyperindex,colorlinks,breaklinks,allcolors=blue]{hyperref}
\usepackage{txfonts}

\usepackage{amsmath, amssymb}
\usepackage{pstricks}
\usepackage{graphicx}
\usepackage{tikz}
\usepackage[utf8x]{inputenc}
\usepackage{color}
\usepackage{bm}

\usepackage{pstool}
\usepackage{pgf}
\usepackage{import}

\usepackage{cleveref}

\usetikzlibrary{snakes}
\usetikzlibrary{arrows}

\usepackage{tabularx}
\usepackage{multirow}
\usepackage{array}
\newcolumntype{L}[1]{>{\raggedright\let\newline\\\arraybackslash\hspace{0pt}}m{#1}}
\newcolumntype{C}[1]{>{\centering\let\newline\\\arraybackslash\hspace{0pt}}m{#1}}
\newcolumntype{R}[1]{>{\raggedleft\let\newline\\\arraybackslash\hspace{0pt}}m{#1}}

\newcommand{\eq}[1]{(\ref{eq:#1})}
\newcommand{\Eq}[1]{Eq.\,\eqref{eq:#1}}

\newcommand{\Fig}[1]{Fig.~\ref{fig:#1}}
\newcommand{\Figs}[2]{Figs.~(\ref{fig:#1})--(\ref{fig:#2})}
\newcommand{\fig}[1]{\ref{fig:#1}}

\newcommand{\Sect}[1]{Sect.~\ref{sec:#1}}

\newcommand{\App}[1]{App.~\ref{app:#1}}

\definecolor{ao(english)}{rgb}{0.0, 0.5, 0.0}
\definecolor{applegreen}{rgb}{0.55, 0.71, 0.0}
\definecolor{cadetblue}{rgb}{0.37, 0.62, 0.63}
\definecolor{cadet}{rgb}{0.33, 0.41, 0.47}
\definecolor{byzantine}{rgb}{0.74, 0.2, 0.64}
\definecolor{orange}{rgb}{1.0, 0.5, 0.0}

\makeatletter
\let\cat@comma@active\@empty
\makeatother

\begin{document}

\title{Dark-Antidark Spinor Solitons in Spin-1 Bose Gases}

\author{C.-M. Schmied}
\email{christian-marcel.schmied@kip.uni-heidelberg.de}
\affiliation{Kirchhoff-Institut f\"ur Physik,
            Ruprecht-Karls-Universit\"at Heidelberg,
             Im~Neuenheimer~Feld~227,
             69120~Heidelberg, Germany}
\author{P. G. Kevrekidis}
\email{kevrekid@math.umass.edu}
\affiliation{Department of Mathematics and Statistics, University of Massachusetts,
Amherst, Massachusetts 01003-4515 USA}

\date{\today}

\begin{abstract}
  We consider a one-dimensional trapped spin-1 Bose gas and numerically explore families
  of its solitonic solutions, namely  antidark-dark-antidark (ADDAD), as well as  dark-antidark-dark (DADD) solitary waves.
  Their existence and stability properties are systematically investigated within the experimentally accessible easy-plane  ferromagnetic phase by means of a continuation over the atom number as well as the quadratic Zeeman energy.
  It is found that ADDADs are substantially more dynamically robust than DADDs. 
  The latter are typically unstable within the examined parameter range. 
  The dynamical evolution of both of these states is explored and the implication of their potential unstable evolution is studied. 
  Some of the relevant observed possibilities involve, e.g., symmetry-breaking instability manifestations for the ADDAD, as well as splitting of the DADD into a right- and a left-moving dark-antidark pair with the anti-darks residing in a different component as compared to prior to the splitting. 
  In the latter case, the structures are seen to disperse upon long-time propagation.
\end{abstract}

\maketitle

\section{Introduction}
\label{sec:Introduction}

Since their experimental realization two-and-a-half
decades ago, Bose-Einstein condensates (BECs) have been of
substantial interest due to their ability to provide
a controllable
playground for exploring macroscopic quantum
phenomena~\cite{PethickSmith,Stringari}. Coherent structures
supported by such weakly interacting gases have played a central
role in the relevant research efforts~\cite{Kevrekidis2015}, sharing many
common features with other fields including nonlinear optics \cite{Kivshar2003}
and water waves~\cite{Ablowitz2011}. Thus, numerous
wave patterns have been studied in BECs, ranging from dark solitons~\cite{Frantzeskakis_2010},
vortices and vortex lines~\cite{Fetter_2001,Fetter_2009}, vortex rings~\cite{Komineas2007} to more complex entities including, e.g., hopfions \cite{Kartashov2014} and potentially long-lived vortex
knots \cite{Barenghi2012,Ticknor2019}.
 On the more practical side,
 some of these excitations such as the dark solitons have been proposed
 as potential qubits with remarkably long lifetimes \cite{Shaukat2017}.

 In addition to the exploration of the most prototypical settings
 involving single-component BECs, in recent years the study of
 multi-component BECs has been of particular interest and has been
 summarized also in recent reviews~\cite{Kevrekidis2015,Kevrekidis2016}.
 Within this setting, the study of genuinely spinorial Bose gases
 has contributed to a wide range of new phenomena since
 its inception~\cite{Stenger1998a}. More specifically, it
 has offered the potential for fractional, as well
 as non-Abelian vortices, for the manifestation
 of spin-textures and transitions between them, for the
 study of spin mixing and numerous other effects involving
 the spin degree of freedom, as summarized, among others,
 in the reviews of~\cite{Kawaguchi2012,Stamper-Kurn2013a}. 
 This is a field of substantial ongoing activity, including, e.g., among other
 recent developments, the observation of universal dynamics
 of spinor gases far from equilibrium~\cite{Prufer:2018hto}. 
 
 Due to their long coherence times, solitonic states are
 of particular interest in these systems. 
 Naturally, the multi-component settings offer numerous possibilities for such states.
 Taking a two-component system with zero background in one of the components,
 dark-bright solitons are well-known solutions~\cite{Kevrekidis2016,busch2001,becker2008}.
 In this setting, one of the components forms a potential well whose absence of atoms invites its filling by atoms of  the second component. 
 If the atoms of the second component lie {\it solely} in that region, we talk about dark-bright solitonic states (assuming that the first component harbors also the phase jump associated with a dark soliton).
 Advancing this idea to three-component systems leads to two prototypical configurations featuring 
 either two dark and one bright or one dark and two bright components. 
 Such states are typically known as DBD (dark-bright-dark) or BDB (bright-dark-bright) solitons.
  The formation of such solitary wave excitations involving the three
 components has been experimentally observed in spin-1 Bose gases~\cite{Bersano2018,Lannig2020}.
 In the latter work, the collisional properties  (i.e., polarization shifts in the vector degree of freedom) of the emerging BDB solitons have been systematically investigated by making use of recent advances enabling a high level of experimental control.
 Very recent theoretical studies showed that DBD and BDB solitons are principal constituents of the phase diagram of non-linear excitations in one-dimensional spin-1 Bose gases \cite{Liu2020, Katsimiga2020b}.

 Somewhat similar states can be found in presence of a
 ground-state-like background in {\it all} components.  
 In case of a two-component system, these states are referred to as dark-antidark solitons~\cite{Frantzeskakis2004}.
 The anti-dark component is characterized by a higher concentration of
 atoms on top of the non-zero background  in the well created by the dark component. 
Such solutions have been theoretically proposed, numerically explored
and experimentally identified in~\cite{Danaila2016}. 
A recent study expanded upon this idea experimentally and theoretically exploring states involving up to six dark-antidark structures~\cite{Katsimiga2020}.
Furthermore, related configurations hinging on the idea of the
complementarity of the components also occur in studies addressing two-species
magnetic solitons in multi-component BECs~\cite{Stringari2016, Chai2019, Farolfi2019}.

 Our aim here is to numerically study the three-component variants of
 dark-antidark states in one-dimensional spin-1 Bose gases.
 In particular, we are interested in structures on top of a ground state
 involving {\it all three} components, a feature critical towards
 formulating anti-dark states. 
 This naturally arises within the experimentally accessible easy-plane ferromagnetic phase of the spin-1 system~\cite{Schmied2020}.
 In analogy to the dark-bright case, we then have two prototypical
 configurations
 involving anti-dark structures:
 On the one hand, it is possible to generate states where two components are of the dark soliton type,
 while only one is anti-dark;
 or, on the other hand, to produce a setting with one dark solitonic component harboring
 two anti-darks in the other two components.
 In each of these cases, the dark soliton(s) play(s) the role of an effective attractive
 potential collecting additional atoms and thus forming a density
 bump (i.e., the antidark solitary wave)
 on top of a finite background in the remaining component(s).
 We label these states as DADD
 (dark-antidark-dark) and  ADDAD (antidark-dark-antidark),
 respectively. 
 Naturally, a state where all three components are dark solitary waves is also present. 
 Yet, given that the latter is more proximal to a single-mode approximation~\cite{Kawaguchi2012,Stamper-Kurn2013a} and
  that here we are interested in anti-dark states, we will not focus on these three-component dark states herein.
  
 We find that the ADDAD state is far more dynamically robust than the DADD which is unstable throughout the examined parameter regime. 
The dynamical breakup of the ADDAD state (when it is unstable) leads to an asymmetric distribution of the anti-darks, involving
 an oscillatory magnetization dynamics.
 The unstable DADD typically splits into a left- and right-moving
 dark-antidark pair in the dynamical evolution in which the
distribution of dark and anti-dark solitary waves among the components is {\it different}
from  the initial one. 
Eventually, the resulting patterns appear to disperse on long timescales in our numerical simulations.
 In addition to the generic variation of the number of atoms, the spinor gas offers the possibility to vary the quadratic Zeeman energy which enables a multi-parametric exploration of the stability of the solitonic structures. 
By means of the respective parameter continuation, we find that the larger the quadratic Zeeman energy and the smaller the atom number, the more dynamically robust the corresponding ADDAD state.

 Our presentation is structured as follows. 
 In \Sect{Model}, we examine the theoretical model.
 In \Sect{NumericalMethods}, we introduce numerical methods to study key features of the ADDAD and DADD states.
 In \Sect{NumericalResults}, we discuss numerical results for both states.
  Finally, in \Sect{Conclusion}, we summarize our findings and
 present our conclusions, as well as a number of directions for future
 studies.

\section{Model}
\label{sec:Model}

In present-day experiments, atoms are typically confined in harmonic trapping potentials.
To reach a quasi one-dimensional regime,  highly anisotropic traps
with longitudinal and transverse trapping frequencies selected to
satisfy the condition $\omega_{\parallel} \ll \omega_{\perp}$ are
used.
While we will focus on such configurations hereafter, we do note that the physical considerations leading to the coherent structures presented in this work are still fully valid in a homogeneous Bose gas. 

For the applicability of our study to experimental systems, we numerically examine the respective one-dimensional model of a spin-1 Bose gas in a highly anisotropic trap.
In this case, the three-dimensional wave functions can be separated into a longitudinal and transverse part.
The transverse wave function, being in the ground state of the respective harmonic oscillator, can then be integrated out to obtain the following system of coupled one-dimensional Gross-Pitaevskii equations (GPEs) for the longitudinal part~\cite{Kawaguchi2012,Stamper-Kurn2013a, Kivshar2007}:

\begin{align}
\label{eq:EOM1}
i \hbar \partial_t \psi_{\pm 1} &= H_0 \psi_{\pm 1} +  q \psi_{\pm 1} +  c_1 \left( \lvert \psi_{\pm 1} \lvert ^2  +\lvert \psi_{0} \lvert ^2 - \lvert \psi_{\mp 1} \lvert ^2\right) \psi_{\pm 1}  \nonumber \\  
& \quad + c_1 \psi_0^2 \psi_{\mp 1}^*  ,  
\end{align}
\begin{equation}
\label{eq:EOM2}
i \hbar \partial_t \psi_{0} = H_0 \psi_{0} + c_1 \left( \lvert \psi_{1} \lvert ^2  + \lvert \psi_{- 1} \lvert ^2\right) \psi_{0} +2 c_1 \psi_{-1} \psi_{ 0}^* \psi_1. 
\end{equation}
Here, $\psi_{\pm 1} \equiv \psi_{\pm 1} (x,t)$ and $\psi_0 \equiv \psi_{0} (x,t)$ are the complex classical bosonic fields that correspond to the magnetic sublevels $m_{\mathrm{F}} = \pm 1, 0$ within the $F = 1$ hyperfine manifold. 
The asterisk denotes complex conjugation. 
The spin-independent part of the Hamiltonian is  $H_0 = - [\hbar^2 /( 2M )] \partial_x^2 + \left(1/2\right) M \omega_{\parallel}^2 x^2 +c_0 {n}_{\mathrm{tot}}$, where $n_{\mathrm{tot}} = \lvert \psi_{1} \lvert ^2  +\lvert \psi_{0} \lvert ^2 + \lvert \psi_{-1} \lvert ^2$ is the total density and $M$ denotes the mass of the atoms.  
Consequently, the total atom number is obtained as $N = \int \mathrm{d}x \, n_{\mathrm{tot}}$.
The parameter $q$ is the quadratic Zeeman energy shift  proportional to an external magnetic field along the $z$-direction. 
It leads to an effective detuning of the $m_{\mathrm{F}} = \pm 1$ components with respect to the
$m_{\mathrm{F}} = 0$ component. 
We have also already absorbed a possible homogeneous linear Zeeman shift in the definition of the fields. 
The parameters $c_0 = c_0^{(3\mathrm{D})}/(2 \pi a_{\perp}^2)$ and $c_1 = c_1^{(3\mathrm{D})}/(2 \pi a_{\perp}^2)$, with $a_{\perp} = \sqrt{\hbar / (M \omega_{\perp})}$ being the transverse harmonic oscillator length, characterize the effectively one-dimensional density-density and spin-spin coupling.  
In the longitudinal direction, the motion in the trap is characterized by  the oscillator length $a_{\parallel} = \sqrt{\hbar /( M \omega_{\parallel})}$.
The above stated three-dimensional coupling constants $c_0^{(3\mathrm{D})}$ and $c_1^{(3\mathrm{D})}$ are given by
\begin{equation}
\label{eq:couplings}
c_0^{(3\mathrm{D})} = \frac{4 \pi \hbar^2 \left(a_0 + 2a_2 \right)}{3M}, \quad c_1^{(3\mathrm{D})} = \frac{4 \pi \hbar^2 \left(a_2 -a_0\right)}{3M},
\end{equation}
in terms of the s-wave scattering lengths $a_0$ and $a_2$. 
In case of $c_1 < 0$, the system is ferromagnetic while for $c_1 > 0$ it is antiferromagnetic.
The characteristic length scale associated with the spin degree of freedom is the spin
healing length $\xi_s = \hbar / \sqrt{2 M n_{\mathrm{tot}} \lvert c_1 \rvert}$, which varies
over the trap due to the inhomogeneity of the density.
The spin healing length typically sets the order of magnitude of the width of three-component solitonic excitations in the spin-1 system.

For our numerical studies, we use experimentally accessible parameters for $^{87}$Rb.
This sets the mass $M$ to the respective rubidium mass.
We take our one-dimensional trap geometry to be characterized by
$(\omega_\parallel , \omega_\perp) = 2 \pi \,  \times  \, (2.5, 250) $ Hz
which is close to the one realized in \cite{Lannig2020}.
The one-dimensional density-density coupling $c_0$ is then inferred from \Eq{couplings} using the s-wave scattering
lengths $a_0 = 101.8 \, a_B$ and  $a_2 = 100.4 \, a_B$ \cite{Kivshar2007}, with Bohr radius $a_B$, and the transverse oscillator length $a_\perp= 0.682 \, \mu m$.
Furthermore, we set the spin-spin coupling to $c_1 = - c_0/ 200$,
 which is in the ballpark of the experimentally relevant value for $^{87}$Rb in the $F=1$ hyperfine manifold.

We obtain a dimensionless form of the equations of motion by rescaling the physical
parameters by a suitable length scale $\ell$. 
This means that $\bar{x} = x/\ell$, $\bar{t}  = t \hbar/(M \ell^2)$, $\bar{\psi}_{\pm1,0}= \psi_{\pm 1,0} \sqrt{\ell}$, $\bar{q} = q M \ell^2/ \hbar^2$, $\bar{c}_{0,1} =  c_{0,1} M \ell / \hbar^2$, $\bar{\omega}_\parallel = \omega_\parallel M \ell^2/\hbar$,
where the overbars denote the dimensionless quantities.
Suppressing the overbar, the dimensionless form of the equations of motion 
is equivalent to \eq{EOM1} and \eq{EOM2} with $\hbar = 1$ 
and $H_0 = - \left(1/2 \right) \partial_x^2 + \left(1/2 \right) \omega_\parallel^2 x^2 +  c_{0} n_{\mathrm{tot}}$. 
 
We perform our numerical simulations on a one-dimensional grid with $N_g = 512$ grid points subject to periodic boundary conditions.
A choice of $\ell =  a_{\parallel} = 6.82 \,  \mu m$ has
  been made to practically facilitate the numerical computations, yet
  our results are reported below in physical units or dimensionless ratios and hence are
  independent  of the concrete selection of $\ell$. 

For a parametric exploration, we vary the total atom number $N$ as well as the quadratic Zeeman energy $q$.
Depending on the quantitative relation between $q$ and the energy associated with the spin interaction, the  
system favors different spin configurations.
This causes the spin-1 Bose gas to feature distinct phases within the plane spanned by the two energy scales.
In order for three-component dark-antidark solitons to exist, we need a non-zero background density in all three 
$m_\mathrm{F}$-components. 
Such a background configuration is generically realized within the easy-plane phase of the spin-1 Bose gas.
In a recent numerical study using the computational
technique of the so-called accelerated continuous-time Nesterov (ACTN) 
scheme, it was found that for a trapped 
spin-1 system in one spatial dimension with vanishing $z$-component of the
magnetization, the system is in the easy-plane phase for quadratic Zeeman energies $q \in (0, 2 n_{\mathrm{p}} \lvert c_1 \rvert)$ in case of $c_1 < 0$ \cite{Schmied2020}.
Here, $n_{\mathrm{p}}$ denotes the peak density of the condensate.
Since the backdrop of the easy-plane phase is critical for the existence of the dark-antidark states of interest, we solely focus on a parametric exploration in the above stated regime of quadratic Zeeman energies.

For the discussion of the magnetic properties of the solitons, recall that the different components of the spin vector in a spin-1 system are given by:
\begin{align}
  F_x &=\frac{1}{\sqrt{2}} \left[ \psi_0 \left(\psi_1^{*} +
    \psi_{-1}^{*}\right) + \left(\psi_1 + \psi_{-1}\right) \psi_0^* \right],
  \label{eq:fx}
\\
  F_y &=  \frac{i}{\sqrt{2}} \left[ \psi_0 \left(\psi_{1}^{*} -
    \psi_{-1}^{*}\right) -  \left(\psi_1 - \psi_{-1}\right) \psi_0^*  \right],
  \label{eq:fy}
  \\
  F_z &= |\psi_1|^2 - |\psi_{-1}|^2,
  \label{eq:fz}
  \\
  F_\perp &= F_x + i F_y = \sqrt{2} \left[ \psi_1 \psi_0^* + \psi_{0} \psi_{-1}^* \right],
  \label{eq:fperp}
\end{align}
where the complex field $F_{\perp}$ denotes the transverse component
of the spin.
Both the integral of the modulus squared of the magnetization vector,
as well as that of the $z$-component of the magnetization are conserved
quantities in the dynamics of Eqs.~(\ref{eq:EOM1}) and \eq{EOM2}.

We remark that we restrict our discussion in this work to the case of ADDAD and DADD states where either the two darks or two anti-darks reside in the $m_\mathrm{F} = \pm 1$ components.
Furthermore, we generally converge to a single soliton solution with a symmetric distribution in the $m_\mathrm{F} = \pm 1$ components implying a vanishing local and global $F_z$ magnetization.

\begin{figure}
  \includegraphics[width= 0.98 \columnwidth]{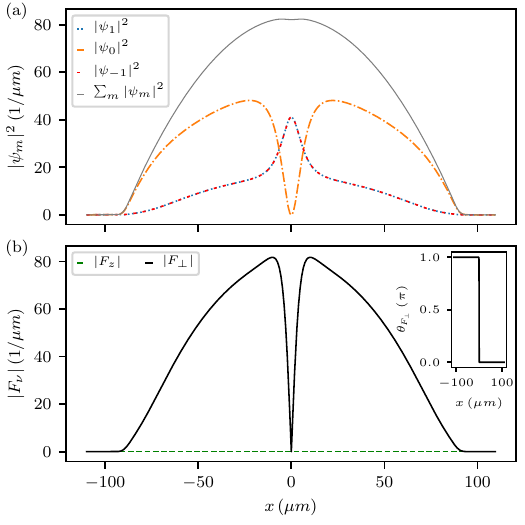}
  \caption{A typical example of an antidark-dark-antidark (ADDAD) state 
    for $N= 10000$ and $q = 0.5 \, n_{\mathrm{p}} \lvert c_1 \rvert$.
    (a) Densities of the three components
    $|\psi_{m}|^2$, $m = \pm  1, 0$, and the total density $\sum_m |\psi_{m}|^2$ (solid grey line).
      The total density shows a small suppression at the position of the ADDAD state.	
     The $m_{\mathrm{F}} = 0$ component carries the dark soliton (dash-dotted orange line). 
    (b) Main frame: Amplitudes of the different components of the magnetization $\lvert F_{\nu}\rvert $, with $\nu = z , \perp$. 
	The ADDAD state has no $F_z$ magnetization (dashed green line), but features a dark soliton in the transversal spin $F_\perp = \lvert F_\perp  \rvert \exp \{ i \, \theta_{F_\perp} \}$ characterized by an amplitude suppression (see solid black line in the main frame) and a  corresponding phase jump (see solid black line
	in the inset). }
\label{fig:ADDADDensSpin}
\end{figure}
\begin{figure}
  \includegraphics[width=0.92 \columnwidth]{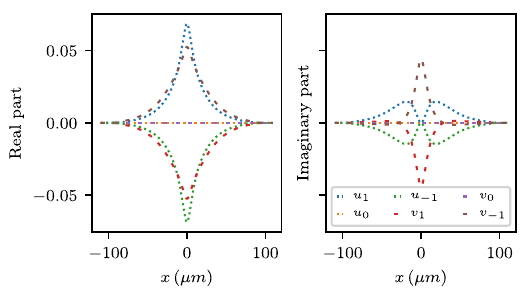}
  \caption{Real and imaginary parts of the normalized mode functions $u_m$, $v_m$ with $m = \pm 1,0$ of the unstable eigendirection of the ADDAD state 
    for $N= 10000$ and $q = 0.5 \, n_{\mathrm{p}} \lvert c_1 \rvert$ as obtained from the numerical evaluation 
    of the respective BdG equations. 
    The corresponding mode frequency is purely imaginary leading to an exponential growth of the depicted unstable eigenmode. 
}
\label{fig:ADDADModeFunc}
\end{figure}

\begin{figure}
  \includegraphics[width=0.92 \columnwidth]{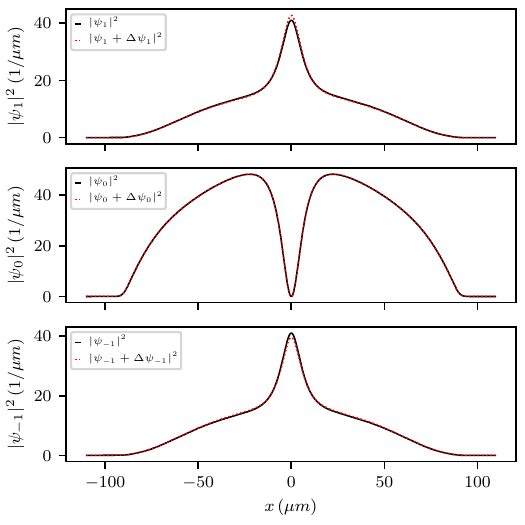}
  \caption{Comparison of the density profiles $\rvert \psi_m \lvert^2$, $m=\pm1,0$ (solid black lines), of the ADDAD state  for $N= 10000$ and $q = 0.5 \, n_{\mathrm{p}} \lvert c_1 \rvert$  with the profiles $\rvert \psi_m + \Delta \psi_m \lvert^2$ (red dotted lines) resulting from adding an exaggerated perturbation of the unstable eigenmode. 
  The perturbation is chosen as $\Delta \psi_m = C (u_m + v_m^*)$ with $C = 20$
  and mode functions $u_m$, $v_m$ as depicted in \Fig{ADDADModeFunc}.
  The unstable eigenmode causes a symmetry breaking between the $m_\mathrm{F} = \pm 1$ components.
}
\label{fig:ADDADPerturbationEffect}
\end{figure}

\section{Numerical Methods}
\label{sec:NumericalMethods}

Our numerical investigation of the ADDAD and DADD states involves three steps:
\begin{enumerate}
\item To find these solitonic structures for a given parameter set within the easy-plane phase of a spin-1 Bose gas,
we employ an exact Newton-Raphson (hereafter referred to, for
simplicity, as just Newton) iterative scheme to the dimensionless, time-independent versions of the equations of motion \eq{EOM1} and \eq{EOM2}, see \App{TIEOM}.
For details on the applied Newton iterative
scheme, see \App{NewtonScheme} and \cite{Schmied2020}.  
To converge to the ADDAD state within the easy-plane phase, we start the Newton iteration 
with the initial ``guess''
\begin{align}
\begin{pmatrix}
\psi_1 \\ \psi_0  \\ \psi_{-1}
\end{pmatrix}
\sim \sqrt{\frac{\mu_{\mathrm{TF}}-V(x) }  {c_0}} \Theta \left( R_{\mathrm{TF}}^2 -x^2 \right)
\begin{pmatrix}
1 \\ \tanh \left[ \frac {x - x_0}{2 \lambda} \right] \\ 1 
\end{pmatrix},
\end{align}
where $x_0$ is the center position of the soliton and $\lambda$ quantifies the width of the hyperbolic tangent. 
The Heaviside theta function $\Theta$ is defined as $\Theta(z) = 1$ for $z \geq 0$.
We place the soliton at the trap center such that we set $x_0 = 0$.
We further take the parameter $\lambda$ to be on the order of the spin healing length $\xi_s $ at the trap center.
We choose a Thomas-Fermi (inverted parabolic) background profile~\cite{Stringari} as we work in a regime
of comparatively large $N$ where the energy associated with the non-linear terms is considerably larger
than the kinetic energy.
Due to the density-density coupling $c_0$ being two orders of magnitude larger than the spin-spin coupling $c_1$, it is reasonable to take a one-dimensional Thomas-Fermi profile characterized by $c_0$, the trapping potential $V(x) =(1/2) M \omega_\parallel^2 x^2$ and the corresponding Thomas-Fermi chemical potential $\mu_{\mathrm{TF}} =(1/2) \left(3 \sqrt{M} c_0 \omega_\parallel N / 2 \right)^{2/3}$. 
The associated Thomas-Fermi radius $R_{\mathrm{TF}}$ is defined via $V(x= R_{\mathrm{TF}})= \mu_{\mathrm{TF}}$.
It is worthwhile to note that near the linear limit, i.e., for a weak nonlinearity (small $N$), the Thomas-Fermi profile has to be replaced by a Gaussian to obtain an adequate initial guess for the Newton iteration scheme. 
For the system studied in this work, we find that the background profile gradually approaches a Gaussian for $N \lesssim 250$; for such small values of $N$, the mean-field description is no longer expected to remain valid, yet it is of mathematical interest in its own right. 
We remark that the Newton iteration converges to the desired solitonic state for both, the Thomas-Fermi and the Gaussian background density profile. 
However, we generally observe that the more accurate the initial
guess, the faster the convergence of the Newton iteration.
Irrespective of the background profile, the presence of the hyperbolic tangent profile in the component bearing the dark soliton will spontaneously play the role of an attractive potential well leading to a mass of atoms in the other two components.

To converge to the DADD state, we start the Newton iteration 
with the initial ``guess"
\begin{align}
\begin{pmatrix}
\psi_1 \\ \psi_0  \\ \psi_{-1}
\end{pmatrix}
\sim \sqrt{\frac{\mu_{\mathrm{TF}}-V(x) }  {c_0}} \Theta \left( R_{\mathrm{TF}}^2 -x^2 \right)
\begin{pmatrix}
\tanh \left[ \frac {x - x_0}{2 \lambda} \right] \\ 1 \\ \tanh \left[ \frac {x - x_0}{2 \lambda} \right] 
\end{pmatrix}.
\end{align}
As indicated also above, converging to the ADDAD or DADD state does not require any knowledge about the form of the anti-dark(s)
in the initial  ``guess" for the Newton scheme.
The crucial part is the presence of a phase jump and an amplitude suppression in the desired dark component(s).

\item Upon converging to the ADDAD and DADD states by means of the
  Newton scheme for a particular set of parameters, we subsequently
  study their stability properties.
These properties are extracted by numerically solving the Bogoliubov de-Gennes (BdG) equations, obtained by considering small perturbations about the solitonic states to linear order. 
To obtain the BdG equations, we take the Ansatz
\begin{equation}
\label{eq:AnsatzPsi}
\psi_m (x,t) = \left[ \Phi_{m}(x) + \epsilon \, \delta \psi_m (x,t) \right] e^{- i \mu t},
\end{equation}
with $m = \pm 1, 0$ labeling the three hyperfine components and
$\Phi_{m}(x)$ being the wave function of each component of the
respective solitonic state obtained from the Newton scheme. Here, $\mu$ is the corresponding chemical potential,
$\epsilon$ is a small parameter with $\epsilon \ll 1$ and $\delta \psi_m$ is the perturbation about the solitonic state.
For the perturbation, we write
\begin{equation}
\label{eq:AnsatzModeFunc}
\delta \psi_m (x,t) =  u_m (x) e^{- i \omega t} + v_m^* (x) e^{i \omega^* t} ,
\end{equation}
with mode frequency $\omega$ and mode functions $u_m , v_m$. 
For the resulting BdG equations and further details on the method see \App{BdG} and \cite{Schmied2020}.
Solving the BdG equations yields the eigenmodes and the respective mode frequencies of excitations about the investigated state.
If all mode frequencies are real, the state is dynamically stable.
Eigenmodes corresponding to mode frequencies with a non-zero imaginary part are dynamically unstable as they grow in time.
Their growth rate $\gamma$ is given by the magnitude of the imaginary part of the mode frequency.
Due to a finite accuracy of the solver used to evaluate the BdG equations, we consider eigenmodes to be unstable only 
when the imaginary part is larger than $10^{-3}$. 
Unstable modes can be of two different kinds.
If the mode frequencies are purely imaginary, we speak of an exponential instability as  the associated mode occupation grows exponentially in time. 
If the mode frequencies are complex, the instability is of oscillatory
nature,
i.e., the growth is accompanied by oscillations. 

\item In the case that the BdG analysis reveals the ADDAD or DADD state to be dynamically 
unstable for a given parameter setting, we aim at investigating the dynamical evolution of the respective state.
We then compute the time evolution of the mean-field model by solving the equations of motion \eq{EOM1} and \eq{EOM2} in dimensionless form by means of a spectral split-step algorithm.
As initial configurations for the respective simulations, we take the ADDAD and DADD states as obtained from the Newton scheme 
and add a small perturbation to the unstable eigendirection(s)
  resulting from the respective BdG analysis in order to seed the instability. 

\end{enumerate}

\section{Numerical results}
\label{sec:NumericalResults}

In this section, we present our numerical results concerning the ADDAD and DADD states.
We start by investigating the key features of the ADDAD states in \Sect{ResultsADDAD} followed
by the DADD states in \Sect{ResultsDADD}.

\begin{figure}
  \includegraphics[width=0.98 \columnwidth]{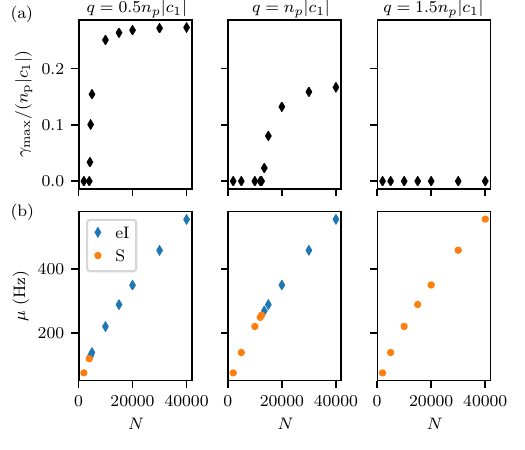} 
  \caption{Stability properties of the ADDAD state for a mono-parametric continuation over the atom number $N$
	for three values of the quadratic Zeeman energy $q =0.5 \, n_{\mathrm{p}} \lvert c_1\rvert$ (left column), $q =n_{\mathrm{p}} \lvert c_1\rvert$ (middle column) and $q =1.5 \, n_{\mathrm{p}} \lvert c_1\rvert$ (right column).  
	(a) Maximal instability growth rates $\gamma_\mathrm{max}$ in units of $n_{\mathrm{p}} \lvert c_1\rvert$.
	The ADDAD structure becomes more stable for larger $q$.
	For $q =1.5 \, n_{\mathrm{p}} \lvert c_1\rvert$, the state is dynamically stable for all
	$N$ considered.
	(b) Bifurcation diagram showing the chemical
    potential $\mu$ as a function of the number of atoms
    $N$. Stability of the ADDAD state is represented by orange
    circles, while
    exponential instability is shown by blue diamonds. 
}
\label{fig:ADDADContinuationOverN}
\end{figure}
\begin{figure}
  \includegraphics[width=0.92 \columnwidth]{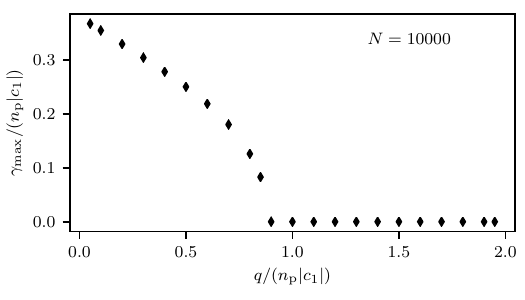}
  \caption{An addition to \Fig{ADDADContinuationOverN}, now showing the
    stability quantified by the maximal instability growth rate $\gamma_{\mathrm{max}}$ as a function of $q$ for a  
    fixed value of $N=10000$ atoms.
    Again, it can be seen that larger $q$'s lead to a reduced instability growth rate and eventually to a complete stabilization of
    the ADDAD structure for $q \gtrsim 0.9 \, n_{\mathrm{p}} \lvert
    c_1\rvert$ for this value of $N$.
    The growth rates and the quadratic Zeeman energies are measured in units of $n_{\mathrm{p}} \lvert c_1\rvert$.
}
\label{fig:ADDADqDependence}
\end{figure}
\begin{figure}
  \includegraphics[width=0.92 \columnwidth]{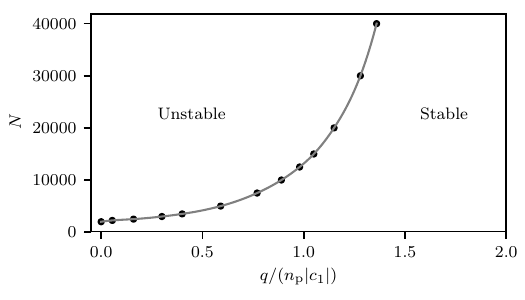}
  \caption{
    Stability of the ADDAD state within the ($q$,$N$)-plane. The black dots mark the boundary between the unstable and stable region of the ADDAD for selected atom numbers. The grey line is a cubic interpolation of the data points.
  The ADDAD becomes more dynamically stable when increasing $q$.   
    The quadratic Zeeman energy $q$ is measured in units of $n_{\mathrm{p}} \lvert c_1\rvert$.
}
\label{fig:ADDADqNplane}
\end{figure}
\begin{figure}
  \includegraphics[width=0.513\columnwidth]{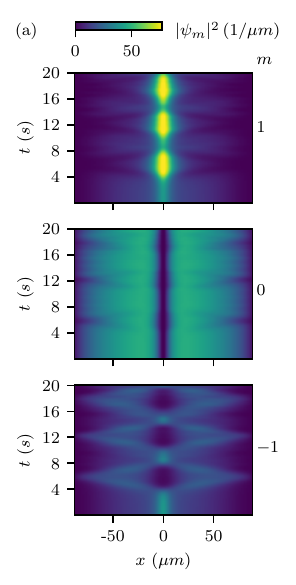}
  \hspace{-0.40cm}
  \includegraphics[width=0.513 \columnwidth]{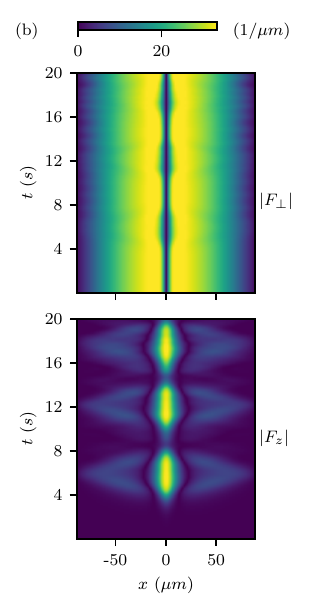}
  \caption{
Time evolution of an unstable ADDAD state for $N= 10000$ and $q = 0.5 \, n_{\mathrm{p}} \lvert c_1\rvert$.
   (a) Densities of the three components $\lvert \psi_m \rvert^2$, $m= \pm 1,0$. 
   The symmetry breaking nature of the destabilizing dynamics can be observed between the
    $m_{\mathrm{F}} = \pm 1 $ components. 
    (b) Amplitudes of the different spin components $\lvert F_\nu \rvert$,
    $\nu = z,\perp$.  
    The persistence of a dark solitonic state in the $m_{\mathrm{F}} = 0$ component is confirmed through the dynamics of $F_\perp$ (top panel), while the asymmetry becomes visible through the nontrivial oscillations of $F_z$ (bottom panel). }
\label{fig:ADDADDynamics}
\end{figure}

\begin{figure}%
  \includegraphics[width= 0.98 \columnwidth]{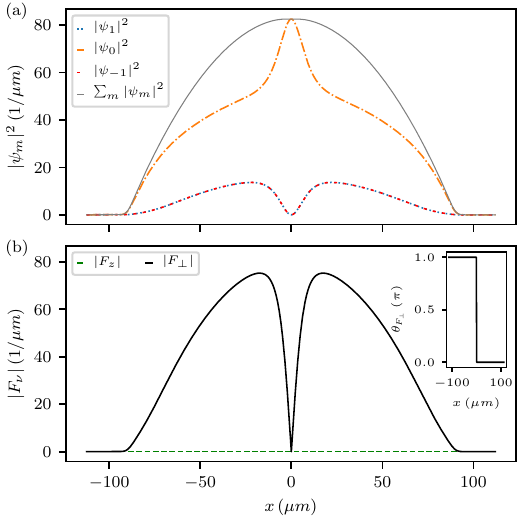}
  \caption{A typical example of a dark-antidark-dark (DADD) state 
    for $N= 10000$ and $q = 0.5 \, n_{\mathrm{p}} \lvert c_1 \rvert$.
    (a) Densities of the three components
    $|\psi_{m}|^2$, $m = \pm  1, 0$, and the total density $\sum_m |\psi_{m}|^2$ (solid grey line).
     The total density shows a tiny suppression at the position of the DADD state.	
     The $m_{\mathrm{F}} = \pm 1$ components carry the dark solitons (dotted blue and dashed red lines). 
    (b) Main frame: Amplitudes of the different components of the magnetization $\lvert F_{\nu}\rvert $, with $\nu = z , \perp$. 
	The DADD state has no $F_z$ magnetization (dashed green line), but features a dark soliton in the transversal spin $F_\perp = \lvert F_\perp  \rvert \exp \{ i \, \theta_{F_\perp} \}$ characterized by an amplitude suppression (see solid black line in the main frame) and a corresponding phase jump (see solid black line in the inset). 
}
\label{fig:DADDDensSpin}
\end{figure}

\begin{figure}
  \includegraphics[width=0.92 \columnwidth]{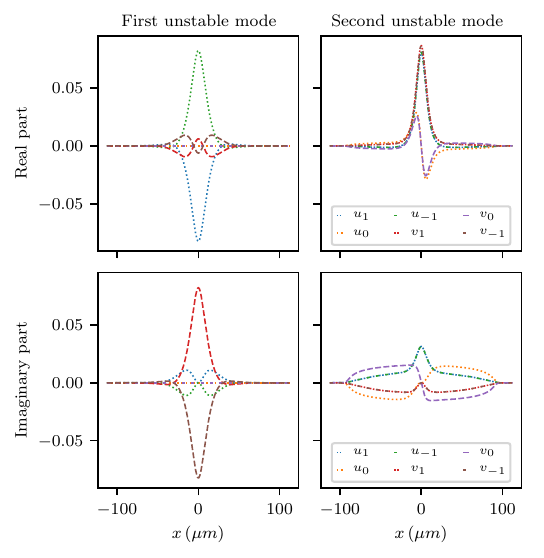}
  \caption{{Real and imaginary parts of the normalized mode functions $u_m$, $v_m$ with $m = \pm 1, 0$ of the two unstable eigendirections of the DADD state for $N= 10000$ and $q = 0.5 \, n_{\mathrm{p}} \lvert c_1 \rvert$ as obtained from the numerical evaluation of the respective BdG equations. 
    The corresponding mode frequencies are purely imaginary leading to an exponential growth of the depicted unstable eigenmodes. 
  The instability growth rate of the first eigenmode (left panel) is significantly larger than for the second eigenmode (right panel).
  Hence, the instability predominantly stems from the most unstable eigendirection.    
    }
}
\label{fig:DADDModeFunc}
\end{figure}
\begin{figure}
  \includegraphics[width=0.92 \columnwidth]{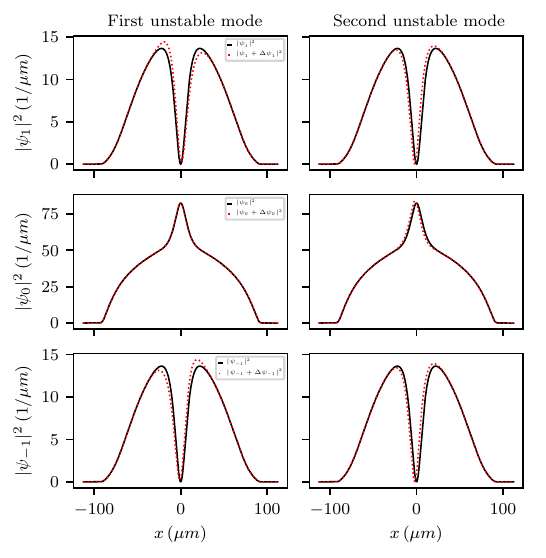}
  \caption{Comparison of the density profiles $\rvert \psi_m \lvert^2$, $m=\pm1,0$ (solid black lines), of the DADD state  for $N= 10000$ and $q = 0.5 \, n_{\mathrm{p}} \lvert c_1 \rvert$  with the profiles $\rvert \psi_m + \Delta \psi_m \lvert^2$ (red dotted lines) resulting from adding exaggerated perturbations of the first (left column) and second (right column) unstable eigenmode. 
  The perturbations are chosen as $\Delta \psi_m = C (u_m + v_m^*)$ with $C = 15$
  and mode functions $u_m$, $v_m$ as depicted in \Fig{DADDModeFunc}.
  The first (most) unstable eigenmode, that will be observed below to
  be dominant in the dynamics,
  breaks the symmetry between the $m_\mathrm{F} = \pm 1$ components
  resulting in an opposite-directed motion of these components. This eventually
  leads to a splitting of the DADD state. The second unstable
  eigenmode is a translational mode as it shifts all three components equally (in the case
  shown here to the left).
}
\label{fig:DADDPerturbationEffect}
\end{figure}
\begin{figure}
  \includegraphics[width=0.98 \columnwidth]{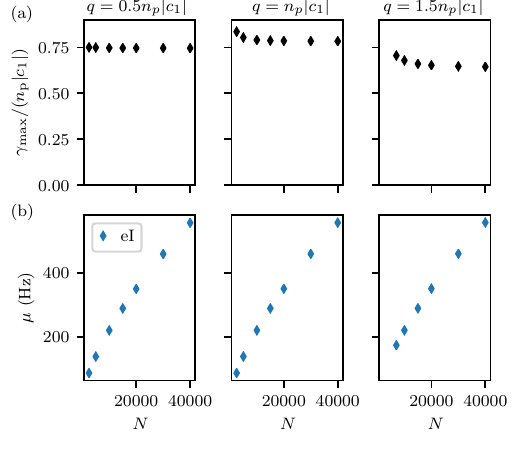} 
  \caption{Stability properties of the DADD state for a mono-parametric continuation over the atom number $N$
	for three values of the quadratic Zeeman energy $q =0.5 \, n_{\mathrm{p}} \lvert c_1\rvert$ (left column), $q =n_{\mathrm{p}} \lvert c_1\rvert$ (middle     
	column) and $q =1.5 \, n_{\mathrm{p}} \lvert c_1\rvert$ (right column).  
	(a) Maximal instability growth rates $\gamma_\mathrm{max}$ in units of $n_{\mathrm{p}} \lvert c_1\rvert$.
	The DADD structure is considerably less stable than the ADDAD state (cf.~\Fig{ADDADContinuationOverN}).
	In particular, it is generically unstable for the
        parameters considered
        with a nontrivial instability growth rate.
	(b) Bifurcation diagram showing the chemical
    potential $\mu$ as a function of the number of atoms $N$.  
    The exponential instability of the DADD state is represented by blue diamonds. 
    Note that for $q =1.5 \, n_{\mathrm{p}} \lvert c_1\rvert$ a minimal atom number of $N \simeq 7000$
    is needed to have a sufficient amount of atoms in the $m_\mathrm{F}= \pm1$ components.
}
\label{fig:DADDContinuationOverN}
\end{figure}
\begin{figure}
  \includegraphics[width=0.92 \columnwidth]{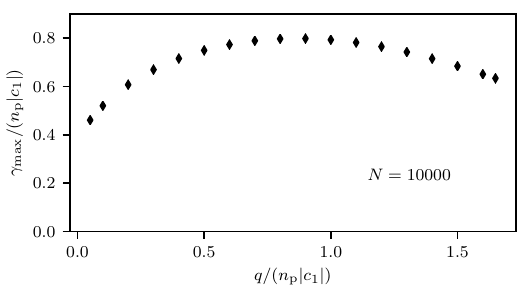}
  \caption{ An addition to \Fig{DADDContinuationOverN}, now showing the
    stability quantified by the maximal instability growth rate $\gamma_{\mathrm{max}}$ as a function of $q$ for a  
    fixed value of $N=10000$ atoms.
    The DADD structure is unstable within the entire parameter regime. 
    The growth rate is largest at $q \simeq 0.9 \, n_{\mathrm{p}} \lvert c_1\rvert$ and drops by a factor of $\simeq 2$ as  
    $q \rightarrow 0$.
    For $q > 1.65 \, n_{\mathrm{p}} \lvert c_1\rvert$ the amount of atoms in the $m_\mathrm{F}= \pm1$ components
    becomes too small to observe a true DADD structure.
    The growth rates and the quadratic Zeeman energies are measured in units of $n_{\mathrm{p}} \lvert c_1\rvert$.
}
\label{fig:DADDqDependence}
\end{figure}

\subsection{ADDAD Solitons}
\label{sec:ResultsADDAD}

A typical example of the ADDAD structure obtained by means of the exact Newton scheme for a total atom 
number of $N=10000$ and a quadratic Zeeman energy of $q = 0.5 \, n_{\mathrm{p}} \lvert c_1\rvert$  is shown in \Fig{ADDADDensSpin}. 
It can be seen that the dark soliton in the $m_{\mathrm{F}} = 0$ component [dashed-dotted line in \Fig{ADDADDensSpin}(a)] creates an effective potential attracting atoms from the two other components [dotted blue and dashed red lines in \Fig{ADDADDensSpin}(a)], and forming the anti-dark spikes in these. 
Given the symmetry of the $m_{\mathrm{F}} = \pm 1$ states, there is no $F_z$ magnetization in this case [see green dashed line in \Fig{ADDADDensSpin}(b)], yet the imprint of the dark soliton leads the transverse component of the magnetization $F_\perp$ to possess also a dark solitonic structure characterized by an amplitude suppression and an associated phase jump [black solid lines in \Fig{ADDADDensSpin}(b)].

Performing the BdG analysis, we find that the potential instability of the ADDAD is characterized by one unstable eigendirection associated with the presence of one dark soliton.
  When an instability is present, the corresponding mode frequency is purely imaginary resulting in an exponential growth of the unstable mode.
From the respective mode functions (see \Fig{ADDADModeFunc}) which are only non-zero for the $m_\mathrm{F} = \pm 1$ components, we infer that the instability acts on the two anti-darks.
The actual effect of the unstable eigendirection can be deduced by comparing the density profiles of the ADDAD state
with the profiles obtained by adding an exaggerated perturbation of the unstable eigenmode to the state (see \Fig{ADDADPerturbationEffect}). 
We find the unstable eigendirection to cause a symmetry breaking between the $m_\mathrm{F} = \pm 1$ components
as the perturbation increases the anti-dark's mass in the $m_\mathrm{F} = 1$ component while it decreases the mass 
in the  $m_\mathrm{F} = -1$ component. 
We will discuss the resulting dynamical evolution of the unstable
ADDAD  state and the corresponding manifestation of this
symmetry breaking effect further below.

The main results of our systematic investigation of the stability of the ADDAD state are summarized in \Figs{ADDADContinuationOverN}{ADDADqNplane}.
\Fig{ADDADContinuationOverN}(a) shows the maximal instability growth rate $\gamma_\mathrm{max}$
as a function of the total atom number  $N$ for three typical values of the quadratic Zeeman energy within the easy-plane phase.
It can be observed that the larger $q$, the wider the interval of stability of the ADDAD state with respect to $N$. 
In fact, for $q=1.5 \, n_\mathrm{p} \lvert c_1 \rvert$, the state is dynamically stable for {\it all} atom numbers considered.
As $q$ is lowered, the stability threshold of the state $N_c$ decreases. 
For $q=0.5 \, n_\mathrm{p} \lvert c_1 \rvert $, stable structures only exist for $N \lesssim 2500$.
\Fig{ADDADContinuationOverN}(b) contains the standard continuation diagram of the chemical potential $\mu$ as a function of $N$ for the different $q$'s representing stability by orange circles and exponential instability by blue diamonds. 
A complementary perspective, fixing the atom number to $N=10000$ and varying $q$, is
presented in \Fig{ADDADqDependence}. 
It can be seen that up to $q \simeq 0.9 \,  n_{\mathrm{p}} \lvert c_1 \rvert$, the structure is unstable for this atom number but as $q$ acquires larger values, the ADDAD state is dynamically stabilized. 
The possible parameter continuation over both, the atom number as well as the quadratic Zeeman energy, further allows
to determine the stable and unstable regions of the ADDAD state within the respective ($q$,$N$)-plane (see \Fig{ADDADqNplane}).
The extracted phase boundary between both regions clearly shows that the ADDAD becomes more dynamically stable
for larger quadratic Zeeman energies and for smaller atom
  numbers. 
Notice that the enhanced stability of the state for larger $q$ can be qualitatively understood by the fact that as $q$ increases, the $m_\mathrm{F}=\pm 1$ anti-dark components are suppressed, ultimately leading to a single dark soliton which is a stable state in the polar phase of the spinor system. 

We now turn to the examination of the dynamical instability of the ADDAD state.
We illustrate a typical
time evolution for $N= 10000$ and $q = 0.5 \, n_\mathrm{p} \lvert c_1\rvert$ in \Fig{ADDADDynamics}. 
Here, we perturb the unstable eigendirection according to the BdG formulation in Eqs.~(\ref{eq:AnsatzPsi}) and \eq{AnsatzModeFunc} such that the amplitude 
of the perturbation is $0.1\%$ of the amplitude of the $m_\mathrm{F} = \pm 1$ components of the ADDAD state obtained from the Newton scheme. 
It can be seen that the ADDAD suffers a symmetry breaking leading to an asymmetric partition of the anti-dark components
as expected from the perturbed density profiles depicted in \Fig{ADDADPerturbationEffect}. 
This means, that one of the two $m_\mathrm{F}=\pm 1$ states  acquires more atoms than the other and subsequently, given the Hamiltonian nature of the model, an oscillatory dynamics ensues between the unstable symmetric state and the presumably more dynamically robust asymmetric state [see \Fig{ADDADDynamics}(a)]. 
During this observed oscillation, the dark solitonic structure
persists in the transversal spin $F_{\perp}$ [see top panel of
\Fig{ADDADDynamics}(b)], while at the same time the asymmetry bestows
a nontrivial oscillatory
dynamics in $F_z$ [see bottom panel of \Fig{ADDADDynamics}(b)]. 
We expect the symmetry breaking to occur when the amplitude of the unstable mode becomes similar to the amplitude of the $m_\mathrm{F} = \pm 1$  components of the initial ADDAD state. 
Hence, an estimate for the corresponding break-up time $t_\mathrm{b} \simeq - (\log A) / \gamma_{\mathrm{max}}= 4 \, s$ can directly be inferred from the respective instability growth rate $\gamma_\mathrm{max} = 0.25 \, n_\mathrm{p} \lvert c_1 \rvert$ and the relative amplitude of  the perturbation $A = 0.001$.
Note that the estimate for $t_\mathrm{b}$ agrees remarkably well with the break-up time observed in the numerical simulation [cf.~\Fig{ADDADDynamics}(a)]. 
Small deviations may arise from additional non-linear effects.

\subsection{DADD Solitons}
\label{sec:ResultsDADD}

Similar features as discussed for the ADDAD structure can be obtained for the DADD soliton. 
A typical example of the DADD structure for a total atom number of $N=10000$ and a quadratic Zeeman energy of $q = 0.5\, n_{\mathrm{p}} \lvert c_1\rvert$ is presented in \Fig{DADDDensSpin}.
It can be seen that the two dark solitons in the $m_{\mathrm{F}} = \pm 1$ components [dotted blue and dashed red lines in \Fig{DADDDensSpin}(a)] create an effective potential attracting atoms from the $m_{\mathrm{F}} =0$ component [dash-dotted orange line in \Fig{DADDDensSpin}(a)], and forming the anti-dark spike in the latter. 
Given the symmetry between the $m_{\mathrm{F}} = \pm 1$ states, there is also no $F_z$ magnetization in this case [green dashed line in \Fig{DADDDensSpin}(b)], yet the imprint of the dark soliton leads the transverse component of the magnetization $F_\perp$ to possess also a dark solitonic structure characterized by an amplitude suppression and an associated phase jump [black solid lines in \Fig{DADDDensSpin}(b)].

Performing the BdG analysis, we find that an instability of the DADD is characterized by two unstable eigendirections associated with the fact that the state features two dark solitons.
The corresponding eigenfrequencies (generically) are purely imaginary resulting in an exponential growth of the unstable 
modes.
However, one of the eigendirections exhibits a significantly larger growth rate $\gamma$ than the other, hence we expect this unstable eigenmode to dominate the instability.
From the respective mode functions  of the most unstable eigendirection (see left column in \Fig{DADDModeFunc}) which are only non-zero for the $m_\mathrm{F} = \pm 1$ components, we infer that the instability acts on the two dark solitons.
As we did previously for the ADDAD state, we compare the density profiles of the DADD state with the profiles obtained by adding an exaggerated perturbation of each of the unstable eigenmodes to the state to deduce the actual effect of the unstable eigendirections (see \Fig{DADDPerturbationEffect}). 
The symmetry breaking nature of the most unstable eigendirection causes the $m_\mathrm{F} = \pm 1$ components to move in opposite directions (see left column in \Fig{DADDPerturbationEffect}).
This induces a splitting of the DADD state in the dynamical evolution which will be discussed further below.
The second unstable eigenmode is a translational mode as it shifts all three $m_\mathrm{F}$-components equally in the same direction (see right column in \Fig{DADDPerturbationEffect}).

The main conclusions of our systematic numerical computations as regards the
dynamical stability of the DADD structures are captured in
Figs.~\ref{fig:DADDContinuationOverN} and \fig{DADDqDependence} and can be summarized as follows.
Generally speaking, DADD structures are considerably {\it less} stable than the ADDADs. 
We suspect that this has to do with the more highly excited nature of the
 DADD state involving more dark solitons than the ADDAD state. 
In fact, this has been illustrated in lower component analyses where it was found that the higher
the number of dark solitons, the more potentially unstable modes exist in the system~\cite{Kevrekidis2015}.
What can be clearly discerned in the different panels of \Fig{DADDContinuationOverN} is that 
no stabilization of the structures is found in our mono-parametric continuation over the number of atoms $N$, for different values of the quadratic Zeeman energy $q$. While for larger values of $N$, the instability appears to slightly weaken, which can be inferred from the lower maximal instability growth rate $\gamma_{\mathrm{max}}$, it does not seem to asymptote towards $\gamma_{\mathrm{max}} \rightarrow  0$ and remains always substantially larger than in the previously discussed case of the ADDADs (cf.~\Fig{ADDADContinuationOverN}).
A complementary perspective, fixing the atom number to $N=10000$ and varying $q$ systematically, is
presented in \Fig{DADDqDependence}. 
It can be clearly seen that the DADD state is generically unstable within the entire easy-plane phase.
In particular, it is found to be most unstable for $q \simeq 0.9 n_\mathrm{p} \lvert c_1 \rvert$. 
Approaching the phase boundaries at $q=2
 n_\mathrm{p} \lvert c_1 \rvert$ and $q=0$, the instability growth rate
  decreases, leading to the observed non-monotonic dependence on $q$.
  In the former case, the configuration passes over to a single-component
  (stable) ground state (of the polar phase), while in the latter one
  it tends to a two-component dark soliton which is more robust than its
  spinor counterpart.

\begin{figure}
  \includegraphics[width=0.513\columnwidth]{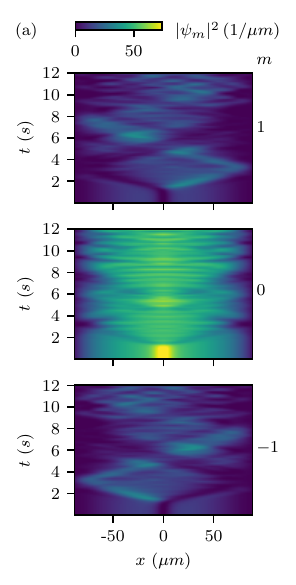}
  \hspace{-0.40cm}
  \includegraphics[width=0.513 \columnwidth]{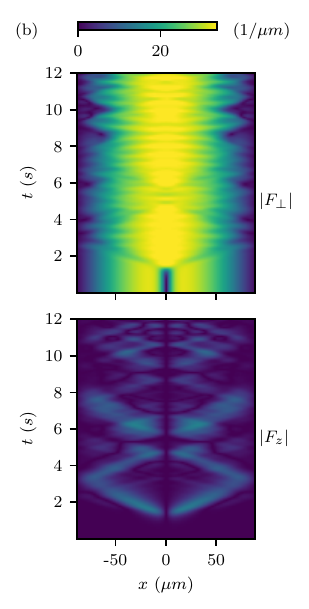}
  \caption{
Time evolution of an unstable DADD state for $N= 10000$ and $q = 0.5 \, n_{\mathrm{p}} \lvert c_1\rvert$.
   (a) Densities of the three components $\lvert \psi_m \rvert^2$, $m= \pm 1,0$. 
        The three panels reveal the definitive splitting of the
         DADD into a left- and a right-moving structure,    with
        the anti-dark being shifted to the $m_\mathrm{F} = 1$ and
        $m_\mathrm{F} = -1$ component, respectively, while the
          $m_\mathrm{F} = 0$ component breaks into a pair of gray
          solitary waves.
        Note that the emerging moving structures are not dynamically
        robust resulting in
        their eventual dispersion over long timescales.
    (b) Amplitudes of the different spin components $\lvert F_\nu \rvert$, $\nu = z,\perp$.  
     In correspondence with (a), we observe a splitting of the $F_\perp$ profile (top panel) concurrently
    with the generation of two oppositely-moving waves
    in the $F_z$ magnetization (bottom panel). 
    After one oscillation in the trap the waves seem to disperse.
}
\label{fig:DADDDynamics}
\end{figure}

\begin{figure}%
  \includegraphics[width= 0.98 \columnwidth]{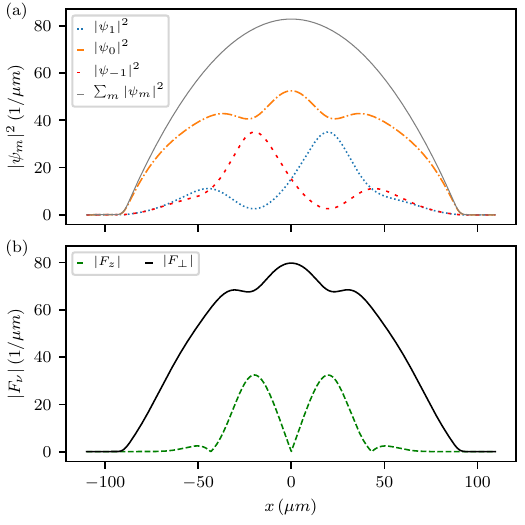}
  \caption{Unstable DADD state shortly after the splitting  (evolution time $t \simeq 1.7 \, s$)
    for $N= 10000$ and $q = 0.5 \, n_{\mathrm{p}} \lvert c_1 \rvert$.
    (a) Densities of the three components
    $|\psi_{m}|^2$, $m = \pm  1, 0$, and the total density $\sum_m |\psi_{m}|^2$ (solid grey line).
     The left- and right-moving gray solitons in the $m_\mathrm{F} = \mp 1$ components are accompanied
     by respective anti-darks in the  $m_\mathrm{F} = \pm 1$ components and shallow gray solitons the $m_\mathrm{F} = 0$
     component.
    (b) Amplitudes of the different components of the magnetization $\lvert F_{\nu}\rvert $, with $\nu = z , \perp$. 
	 Due to the asymmetry between the $m_\mathrm{F} = \pm 1$ components in the left- and right-moving structures, we observe
	 a non-zero local $F_z$ magnetization (dashed green line).
}
\label{fig:DADDSplittingDensSpin}
\end{figure}

The overwhelmingly more unstable nature of the DADD structures naturally
raises the question of the dynamical implications of this instability. This is clarified
in the dynamical evolution of the DADD state for a prototypical
case with
$N= 10000$ and $q = 0.5 \, n_\mathrm{p} \lvert c_1\rvert$ depicted in
\Fig{DADDDynamics}.
Here, we perturb the two unstable eigendirections according to the BdG formulation in Eqs.~(\ref{eq:AnsatzPsi}) and \eq{AnsatzModeFunc} such that the amplitude 
of the perturbation is $0.1\%$ of the amplitude of the $m_\mathrm{F} = \pm 1$ components of the DADD state obtained from the Newton scheme.
In as far as we have been able to observe in our numerical simulations,
the time evolution shown is representative for the relevant dynamics. 
From the perturbed density profiles in \Fig{DADDPerturbationEffect}
  we already observed that the instability eigenvector  corresponding to the most unstable eigendirection breaks the symmetry between the two dark solitons in the $m_\mathrm{F}=\pm 1$ components inducing an opposite-directed motion.
This leads to a splitting of the DADD [see \Fig{DADDDynamics}(a) and \Fig{DADDSplittingDensSpin} for a slice of the density and spin profiles at $t \simeq 1.7 \, s$, i.e., shortly after the splitting] whereby for example the dark soliton in the $m_{\mathrm{F}} = -1$ component captures part of the mass from the anti-dark in the $m_{\mathrm{F}}= 0$ component such that it becomes gray and moves to the left.
Within this process, another fraction of the $m_{\mathrm{F}}=
0$-state's anti-dark mass is captured by the $m_{\mathrm{F}}= 1$
component where an anti-dark soliton emerges.
As a consequence, the  corresponding $m_{\mathrm{F}}= 0$ component is found to feature
an amplitude suppression in the form of a gray solitary wave after the splitting. 
The entire structure appears to be again of DADD type with the
anti-dark being shifted to the $m_{\mathrm{F}}= 1$ component while a
gray solitary wave is present in the $m_{\mathrm{F}}= 0$ component.
Analogously, the right-moving structure
is also of DADD
type with the anti-dark being shifted to the $m_{\mathrm{F}}= -1$
component and another gray soliton arising in the $m_{\mathrm{F}}= 0$ component.
Again, we expect the splitting to occur when the amplitude of the most unstable mode becomes similar to the amplitude of the $m_\mathrm{F} = \pm 1$  components of the initial DADD state. 
Accordingly, an estimate for the break-up time $t_\mathrm{b} \simeq - (\log A) / \gamma_{\mathrm{max}}= 1.33 \, s$ can directly be inferred from the respective instability growth rate $\gamma_\mathrm{max} = 0.75 \, n_\mathrm{p} \lvert c1 \rvert $ and the relative amplitude of  the perturbation $A = 0.001$. 
Note that once again the estimate for $t_\mathrm{b}$ agrees remarkably well with the break-up time observed in the numerical simulation [cf.~\Fig{DADDDynamics}(a)].  
We remark that the structures emerging after the splitting of the DADD are not dynamically robust and quickly generate additional excitations in the system
leading to an eventual dispersion over long timescales. 
We have observed this type of splitting and the creation of moving
solitary waves in the entire range of $q$ values depicted in \Fig{DADDqDependence}.
We find the emerging states to become more robust as $q \rightarrow 0$.
We attribute this property to the increasing fraction of atoms in the $m_{\mathrm{F}} = \pm1$  components  
with respect to the $m_{\mathrm{F}} =0$ component in the initial DADD structure.

\section{Conclusion and outlook}
\label{sec:Conclusion}

In this work, we numerically showed the existence of DADD and ADDAD solitonic states within the 
easy-plane phase of a ferromagnetic trapped one-dimensional spin-1 Bose gas.
Furthermore, we investigated their stability properties by solving the respective BdG equations.
In particular, we elaborated on the stability of the states for a continuation over
the total atom number and the quadratic Zeeman energy.
The identified potential dynamical instabilities of the states were
complemented
with direct numerical simulations elucidating the corresponding dynamics.

Our key observation is that the ADDAD structure is more robust than the DADD state which
is found to be unstable in  all parameter regimes considered.
We suspect that these findings stem from the more highly excited nature of the DADD state
bearing two dark solitons rather than one.
Studying the time evolution of an unstable DADD state revealed that it breaks up into a left- and right-moving pair of
dark-antidark states accompanied by a redistribution of the solitary
waves between the components and the eventual dispersion of the resulting
structures.
However, the evolution of an unstable ADDAD state generated an asymmetric distribution of the anti-darks involving
an oscillatory dynamics of the $F_z$ magnetization. 
Nevertheless, the ADDAD structure could be stabilized in suitable regimes in parameter space and hence be accessible to the recent experimental observations in spinor systems~\cite{Bersano2018,Lannig2020}.

An interesting future direction would  be to generate multiple such
solitary waves and examine in their interactions as it was done recently for BDB solitons in~\cite{Lannig2020}. 
Going one step further, it may be possible to engineer soliton lattices or random soliton gases \cite{Mitschke1997,Schmidt2012,Kevrekidis2015b,Mordant2019} of these three-component structures and study their collisional properties in the spin degree of freedom.
Lastly, it appears especially relevant to
consider generalizations of such states to higher-dimensional
configurations, involving topologically charged coherent structures.
In addition to a scenario of two vortices trapping an anti-dark wave
and a single vortex trapping two anti-darks, further possibilities emerge
given that the multiple vortices can bear the same or
opposite topological charge.
Additionally, both two-dimensional and three-dimensional extensions of linear, planar, or spherical~\cite{Kevrekidis2015, Frantzeskakis_2010} dark solitons trapping corresponding anti-dark states
may be accessible and quite interesting for future work, especially given their potential transverse instabilities and associated dynamics.


\textit{Acknowledgments.} 
The authors thank T.~Gasenzer for his valuable comments that helped us to significantly improve our manuscript
and M.~K.~Oberthaler for discussions and collaboration on the topics described here. 
P.~G.~K.~additionally acknowledges discussions with S.~I.~Mistakidis and G.~C.~Katsimiga
on spinor condensates.
This work was supported by the Heidelberg Graduate School of
Fundamental Physics (HGSFP), the Heidelberg Center for Quantum
Dynamics (CQD), the European Commission, within the Horizon-2020
programme, through the ERC Advanced Grant EntangleGen (Project-ID
694561) as well as the Deutsche Forschungsgemeinschaft (DFG, German Research Foundation) 
-- Project-ID 273811115 -- SFB 1225.
This material is based upon work supported by the US National Science
Foundation under Grants No. PHY-1602994 and DMS-1809074 and
by the Alexander von Humboldt Foundation. 
P.~G.~K.~further acknowledges support from the Leverhulme Trust via a
Visiting Fellowship and thanks the Mathematical Institute of the University
of Oxford for its hospitality during part of this work.

\begin{appendix}
\begin{center}
\textbf{APPENDIX}
\end{center}
\setcounter{equation}{0}
\setcounter{table}{0}
\makeatletter

In the appendix, we briefly elaborate on the numerical methods used to show the existence and to study
the stability properties of the ADDAD and DADD states.
The following discussion is partly taken and adapted from \cite{Schmied2020}, where the interested reader
can also find additional details on the numerical schemes.

\section{Time-independent equations of motion}
\label{app:TIEOM}

In this work, we are interested in ADDAD and DADD solitons that are stationary states of the equations of motion \eq{EOM1} and \eq{EOM2}.
Hence, we aim to identify solutions to the time-independent version of these equations.
By choosing the Ansatz $\psi_m (x,t) = \psi_m (x) e^{-i \mu_m t}$ with $m = \pm 1,0$ and $\mu_m$ being the chemical potential of each spinor component, a  stationary state resulting from Eqs.~(\ref{eq:EOM1}) and \eq{EOM2} has to fulfill the phase matching condition $2 \mu_0 - \mu_1 - \mu_{-1} = 0$.   
As a population imbalance between the $m_{\mathrm{F}} = \pm 1$ components is not favored, independent of the choice of the couplings in the equations of motion, we assume that $\mu_1 = \mu_{-1}$ here, which implies that $\mu_0 = \mu_1 = \mu_{-1}  \equiv \mu$. 
The time-independent equations of motion (in dimensionless form according to the definitions in the main text) then read:
\begin{align}
\label{eq:EOM1TI}
\mathcal{F}_{\pm 1} (\psi_1, \psi_0, \psi_{-1}, \psi_1^*,  \psi_0^*,  \psi_{-1}^*) \equiv &- \mu \psi_{\pm 1} + H_0 \psi_{\pm 1} +  q \psi_{\pm 1}   \nonumber \\ & +  c_1 \left( \lvert \psi_{\pm 1} \lvert ^2  +\lvert \psi_{0} \lvert ^2 - \lvert \psi_{\mp 1} \lvert ^2\right) \psi_{\pm 1}  \nonumber \\ &+   c_1 \psi_0^2 \psi_{\mp 1}^* \nonumber \\ =& \, 0 ,  
\end{align}
\begin{align}
\label{eq:EOM2TI}
\mathcal{F}_0  (\psi_1, \psi_0, \psi_{-1}, \psi_1^*,  \psi_0^*,  \psi_{-1}^*) \equiv& -\mu \psi_{0} + H_0 \psi_{0}
\nonumber \\&+ c_1\left( \lvert \psi_{1} \lvert ^2  + \lvert \psi_{-1} \lvert ^2\right) \psi_{0}  \nonumber \\ &+2  c_1  \psi_{-1} \psi_{ 0}^* \psi_1 \nonumber \\ =& \, 0. 
\end{align}
Here, we introduced functions $\mathcal{F}_{0, \pm 1}$ as abbreviations for the time-independent equations of motion which will be of practical use for discussing the Newton scheme (see \App{NewtonScheme}) as well as the Bogoliubov-de Gennes equations (see \App{BdG}). 

\section{Iterative Newton scheme}
\label{app:NewtonScheme}

Various first- and second-order methods can be applied to find solutions to Eqs.~(\ref{eq:EOM1TI}) and \eq{EOM2TI}. 
Here, we make use of a Newton iteration scheme.
It is a second-order method  which involves the explicit calculation of the Jacobian.
The Newton scheme is not restricted to finding ground states (i.e.~the global energy minimum) of a physical system
such that in case of an adequate initial ``guess'' for the wavefunctions, it offers the possibility to converge to the 
desired ADDAD and DADD states. 

The Newton scheme for the spin-1 system can be cast into the form of a six-dimensional matrix equation:
\begin{equation}
\label{eq:NewtonScheme}
J \Delta \psi = \mathcal{F},
\end{equation}
where $\Delta \psi$ gives the correction to the wave function of the previous iteration of the Newton scheme with $\psi = ( \psi_1,\psi_0,\psi_{-1}, \psi_1^*, \psi_0^* , \psi_{-1}^* )^T$ being a vector of all spinor fields. The vector \kern 0.2em $\mathcal{F} = ( \mathcal{F}_1,\mathcal{F}_0,\mathcal{F}_{-1},\mathcal{F}_1^* ,\mathcal{F}_0^*,\mathcal{F}_{-1}^*)^T $ contains the time-independent equations of motion [see Eqs.~(\ref{eq:EOM1TI}) and (\ref{eq:EOM2TI})] as well as their complex conjugated versions.  
The Jacobian $J$ is given by the matrix
\begin{equation}
\label{eq:Jacobian}
J_{ij} = \frac {\partial \mathcal{F}_i} {\partial \psi_j},
\end{equation}
where $i,j \, \epsilon \,  \{0, \dots,5\}$ and the partial derivative is evaluated at the wave function $\psi$ of the current iteration step.

To converge to a state with fixed atom number $N$, we further introduce a Lagrange multiplier $\lambda$ for the chemical potential $\mu$. 
This adds the following constraint to our Newton scheme
\begin{equation}
\mathcal{F}_{\lambda} \equiv  \int \left ( \lvert \psi_1 \rvert^2 +\lvert \psi_0 \rvert^2 + \lvert \psi_{-1} \rvert^2 \right) \mathrm{d}x - N = 0.
\end{equation}
Consequently, the resulting modified scheme can be written as
\begin{equation}
\label{eq:NewtonSchemeWithLagrangeMult}
\tilde{J} \Delta \tilde{\psi} = \tilde{\mathcal{F}},
\end{equation}
with $\tilde{\psi} = ( \psi_1,\psi_0,\psi_{-1}, \psi_1^*, \psi_0^* , \psi_{-1}^*, \lambda )^T$ and $\tilde{\mathcal{F}} = ( \mathcal{F}_1,\mathcal{F}_0,\mathcal{F}_{-1},\mathcal{F}_1^* ,\mathcal{F}_0^*,\mathcal{F}_{-1}^*, \mathcal{F}_\lambda)^T$. 
In each iteration step we calculate $\tilde{\mathcal{F}}$ and evaluate the corresponding Jacobian $\tilde{J}$ of the system.
The second derivative occurring in the equations of motion is obtained by means of a second-order center difference scheme.
By solving the eigenvalue equation \eq{NewtonSchemeWithLagrangeMult}, we obtain the correction to the wave function $\Delta \tilde{\psi}$. 
The Newton scheme terminates if the norm of the correction is smaller than the preset tolerance of $10^{-10}$.

\section{Bogoliubov de-Gennes equations}
\label{app:BdG}

The stability properties of the ADDAD and DADD state are deduced from numerically solving the corresponding Bogoliubov de-Gennes (BdG) equations.

To derive the BdG equations, we consider a small perturbation about the stationary state of interest. Therefore, we take the Ansatz
\begin{equation}
\label{eq:AnsatzPsiApp}
\psi_m (x,t) = \left[ \Phi_{m}(x) + \epsilon \, \delta \psi_m (x,t) \right] e^{- i \mu t},
\end{equation}
with $m = \pm 1,0$ labeling the three hyperfine components and $\Phi_{m}(x)$ being the wave function of each component at the stationary state; $\mu$ is the corresponding chemical potential; 
$\epsilon$ is a small parameter with $\epsilon \ll 1$ and $\delta \psi_m$ is the perturbation about the stationary state. 

Plugging the Ansatz \eq{AnsatzPsiApp} into  the equations of motion \eq{EOM1} and \eq{EOM2} (in dimensionless form according to the definitions in the main text) and subsequently linearizing the resulting equations (i.e.~taking contributions to order $\epsilon$) yields
\begin{align}
\label{eq:EOMOrderEpsilon}
	 i  \partial_t \delta \psi_m =& \,\,\left(\frac {\partial   \mathcal{F}_m }{\partial \Phi_1}\right)_{| \Phi} \delta \psi_1  +    \left( \frac {\partial   \mathcal{F}_m }{\partial \Phi_0} \right)_{| \Phi}  \delta \psi_0  +  \left(\frac {\partial   \mathcal{F}_m }{\partial \Phi_{-1}} \right)_{| \Phi}  \delta \psi_{-1}    \nonumber \\
	&+  \left( \frac {\partial   \mathcal{F}_m}{\partial \Phi_1^*} \right)_{| \Phi}  \delta \psi_1^*  +  \left(\frac {\partial   \mathcal{F}_m }{\partial \Phi_0^*} \right)_{| \Phi}  \delta \psi_0^*  +  \left(\frac {\partial   \mathcal{F}_m }{\partial \Phi_{-1}^*} \right)_{| \Phi}  \delta \psi_{-1}^* .
\end{align}
Here, the $\mathcal{F}_m$ are the functions introduced in Eqs.~(\ref{eq:EOM1TI}) and \eq{EOM2TI}. 
The partial derivatives of $\mathcal{F}_m$ are taken with respect to the stationary fields and are then evaluated at $\Phi$, with
$\Phi = (\Phi_1 ,\Phi_0 ,\Phi_{-1} ,\Phi_1^* ,\Phi_0^* ,\Phi_{-1}^* )$ being a vector that contains the wave functions at the stationary state. 

To solve the BdG equations, we make use of the Ansatz
\begin{equation}
\label{eq:AnsatzModeFuncApp}
\delta \psi_m (x,t) = \left( u_m (x) e^{- i \omega t} + v_m^* (x) e^{i \omega^* t} \right),
\end{equation}
with mode frequency $\omega$ and mode functions $u_m , v_m$. 
Inserting the Ansatz into \Eq{EOMOrderEpsilon} and matching the phase factors to obtain a time-independent description, we can write the BdG equations as an eigenvalue problem of the form
\begin{equation}
\label{eq:BdGEquations}
\bar{J} \mathcal{M} = - \omega \mathcal{M}.
\end{equation}
Here, $\mathcal{M} = (u_1,u_0,u_{-1},v_1,v_0,v_{-1} )^T$ is a vector that contains all eigenmodes of the system. 
The matrix $\bar{J}$ turns out to be the Jacobian introduced in Eq.~(\ref{eq:Jacobian}) whose lower
half of entries is multiplied by a factor of $-1$. 
Formally, we can write it as
\begin{equation}
\bar{J}_{ij} = \left[1 -2 \Theta(i-3) \right] \left(J_{ij} \right)_{| \Phi} ,
\end{equation}
where $i,j \, \epsilon \, \{0,\dots,5\}$ and the Heaviside theta function $\Theta$ is defined as $\Theta (z) = 1$ for $z \geq 0$.

The mode frequencies $\omega$ correspond to the eigenvalues of $\bar{J}$ and the mode functions $u_m, v_m$ are given by the eigenvectors.
We numerically solve the eigenvalue problem in Eq.~(\ref{eq:BdGEquations}) using the standard $_{-}$geev LAPACK routines in python.

\end{appendix}


%

\end{document}